\documentclass[summary]{URSIGASS2020}

\usepackage{amssymb}
\usepackage{amsmath}

\usepackage[numbers]{natbib}	

\title{COMPASS: VLBI Beacons In Support of Lunar Science and Exploration}

\author{T. M. Eubanks}

\affiliation{%
{Space Initiatives Inc\\
Palm Bay, Florida, USA\\
tme@space-initiatives.com\\}
}

\begin{document}

\maketitle

\begin{abstract}

The large constellations of spacecraft planned for use in cislunar space (on the Lunar surface, in Lunar orbit, and in the vicinity of the Lunar Gateway) require new solutions for positioning, navigation and timing (PNT). 
Here, I describe
 COMPASS (Combined Observational Methods for Positional Awareness in the Solar System), a spacecraft navigation system 
 to provide cost-effective techniques for the positioning of large numbers of spacecraft in cislunar space. 
 COMPASS will use beacons that emit coherent ultra-wideband signals 
designed to be interoperable with existing and future Very Long Baseline Interferometry (VLBI) networks. 
Using differential VLBI, COMPASS will provide rapid determination of the interferometric phase delay 
with picosecond level accuracy during routine VLBI observing sessions. Multi-baseline phase-referenced COMPASS-VLBI observations with simultaneous calibrator observations should thus enable sub-meter accuracy transverse 
positioning 
and meter level lunar orbit determination using 
with small femtospacecraft beacons and a few seconds of observation per position determination. 

\end{abstract}

\section{Introduction}

Some of the earliest VLBI observations of any sort were done using the 
 Apollo Lunar Surface Experiments Package (ALSEP) transmitters, which had very narrow bandwidths of a few kHz
at S-band (the network was spread in frequency from 2275.5 to 2279.5 MHz). Initial experiments without any calibrator sources were used to determine relative positions of the ALSEP transmitters on the Moon to within ``about 30 m along the earth-moon direction and about 10 m in each of the two transverse coordinates'' \cite{King-et-al-1976-a}.
Later work developed differential VLBI including quasar calibrator sources \cite{Slade-et-al-1977-a}, but  unfortunately this work was terminated when the program was canceled. 

Although there was subsequently intensive development of spacecraft navigation using VLBI, incuding tracking of both Cassini and Huygens at Saturn and Titan (see, e.g., \citep{Pogrebenko-et-al-2004-a,Jones-et-al-2014-a,Cimo-et-al-2015-a,Duev-et-al-2016-a}) VLBI was not applied again in lunar operations until 2007, when the Japanese Selene spacecraft carried two small subsatellites into lunar orbit. These small spacecraft, Rstar and Vstar, were used in the VRAD (differential VLBI RADio sources) mission in order to improve the knowledge of the lunar gravity field, particularly on the far side \cite{Goossens-et-al-2011-a}. VRAD, using same-beam differential VLBI (where all radio sources are simultaneously observabled by each radio antenna), demonstrated $<$ 1 picosecond differential phase delay errors. The VRAD VLBI data were used operationally to decrease the VRAD orbit errors from 100s of meters to about 10 meters \cite{Kikuchi-et-al-2009-a} and improve the resulting gravity field.

The Chang'e 3 lunar lander landed on the Moon on 14 December, 2013, as part of Chinese Lunar Exploration Program (CLEP). VLBI was used to track the Chinese Chang'e 3 lander during its landing sequence \cite{Klopotek-et-al-2019-a}  and after landing; in addition VLBI was used to determine rover-lander relative positions on the lunar surface \cite{Liu-et-al-2014-a,Yan-et-al-2011-a}.
The relative positions of the rover were determined at the meter level and the absolute position of the lander to within about 10 meters. 
Chang'e 3 was also observed in geodetic VLBI experiments under the IVS 
OCEL (Observing Chang'e-3 with VLBI)
Research and Development project 
  \cite{Zhang-et-al-2017-b}. Note that the OCEL observations were inserted into conventional geodetic observing sessions and did not hamper their geodetic use.  
  
\section{Extending the Space Service Volume to the Moon}
\label{sec:SSV-Limits} 

The current Space Service Volume (SSV)
extends from 3000 to 36,000 km altitude, \cite{Bauer-et-al-2006-a}, denoting the region near and above the GPS and 
GNSS constellations where it is typically not possible to perform near-instantaneous GNSS point solutions, but instead the GNSS pseudorange observable must be used in a integrated orbital solution. 
Ashman \textit{et al.}
\cite{Ashman-et-al-2018-a} invested GPS performance above the current SSV. At altitudes $>$ 36,000 km all GPS observation is ``over the shoulder'' (i.e., observing satellites on the other side of the Earth) and the use of the GPS signal sidebands is essential to routine GPS availability, making up about 80\% of the signals received at the highest altitudes. 
Simulation of navigation system errors on lunar trajectories exhibited 
considerable Geometrical Dilution of Precision (GDOP), km level radial errors, 100 m lateral errors, and high correlations between the radial position and the spacecraft clock offset. 

A continuously available lunar beacon service providing one-way autonomous tracking would help to extend the SSV to cislunar space and the lunar surface; especially once a lunar communication relay system is available. There is no intrinsic reason why COMPASS position fixes could not processed in a matter of minutes and relayed immediately, if needed, to the spacecraft operators or to the spacecraft itself.

\begin{table}[!t]
\renewcommand{\arraystretch}{1.0}
\caption{The Beamwidth  at the Moon from the Earth for possible antenna and frequency choices. All beamwidths are 
significantly less than the lunar radius of $\sim$1738 km; differential phase VLBI will thus benefit from multiple antennas at each location or
geodetic reference beacons at the Moon.}
\label{table:beam-widths}
\centering
\begin{tabular}{ c c c c c c }
Network & Antenna & \multicolumn{3}{c}{Frequency}\\ 
        &  Diameter & 8.4 GHz & 13 GHz & 43 GHz \\ 
\hline
VLBA & 25 m & 550 km & 360 km & 110 km\\
NG-VLA & 18 m & 760 km & 500 km & 150 km \\
VGOS   & 13 m & 1060 km & 680 km & - \\ 
\hline
\end{tabular}
\end{table}

\section{Tracking Lunar Spacecraft with Very Long Baseline Interferometery}
\label{sec:Tracking}

The basic equation for relatively narrow band coherent VLBI SNR is \cite{Pogrebenko-et-al-2004-a} 
\begin{equation}
SNR = t_{int} \frac{\eta\ \pi\ D^{2}}{4\ k T_{sys}} \left( \frac {P\ G_{transmit}}{4\ \pi\ f\ R^2} \right)
\label{eq:SNR}
\end{equation}
where $\eta$ is the net VLBI receive antenna efficiency, k the Boltzmann constant, T$_{sys}$ the system temperature, D the antenna diameter, R the distance to the source, P the transmitter power, f is the UWB loss factor, 
about 11.3 for 802.15.4a, and G$_{transmit}$ the beacon gain (assumed to be $\sim$1). 
Assuming typical values for the 25 meter VLBA telescopes, with 
T$_{sys}$ = 26.6 K and $\eta$ = 0.55,
a 1 milliWatt UWB transmitter at the lunar distance could in theory be detected with an SNR of 10
with 1 second integration. On the other hand, radio observers without knowledge or interest in the UWB signal structure, i.e., those treating it as 
an incoherent broadband source would, at the lunar distance, observe a broadband signal power of 0.4 milliJanskies, well below the sensitivity of most terrestrial users (and with many natural radio sources with similar or greater power in the sky). 

With signal multiplexing enabling multiple beacons to broadcast simultaneously, the
existing and planned VLBI networks could potentially track thousands of spacecraft  carrying suitable VLBI beacons, providing a new source of financial support for VLBI. Terrestrial VLBI observations can provide 
differential phase data with picosecond accuracy \citep{Kikuchi-et-al-2009-a}, corresponding  
(with a 1.5 Earth radius baseline) to an instantaneous transverse accuracy of $\sim$3 $\times$ 10$^{-11}$ radians
or $\sim$7 $\mu$as. That would produce a transverse positioning error of $\sim$1 cm and a radial (parallax) error of 
$\sim$0.5 meters, assuming observations from a suitably distributed network of VLBI stations on the Earth. 
(Note that near the lunar limb, e.g., in the polar regions,  ``transverse'' and ``radial'' will be significantly mis-aligned with the Lunar topography.)

\section{COMPASS}

\begin{figure}
\centering
\includegraphics[scale=0.25]{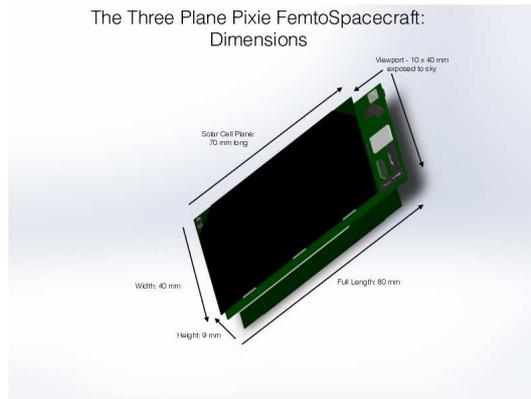}
\caption{The AIM/FAIM Pixie Asteroid Femtospacecraft (without insulation), from \cite{Eubanks-et-al-2017-a}.}
\label{fig:Pixie-2}
\end{figure}

COMPASS was first proposed (under a different name) as part of FAIM, the Femtospacecraft Asteroid Impact Mission \citep{Eubanks-et-al-2017-a}, a proposal to provide instruments and VLBI beacons for the asteroid moon Didymos-$\beta$ for its use in a planetary defense experiment. 
The Moon is much closer than the Didymos impact test, and it is thus correspondingly easier to provide a useful VLBI signal from a small module. The full scale COMPASS module shown in Figure \ref{fig:Pixie-2} would be (at least during the lunar day)  able to provide rapid determination of surface positions or lunar orbiters
using multi-baseline phase-referenced COMPASS VLBI observations with simultaneous or near simultaneous calibrator observations. The module in Figure \ref{fig:Pixie-2} would be able to act autonomously; 
smaller (credit-card sized) beacons could also provide useful data on a pre-programmed transmission schedule.

Low power ($\lesssim$ 10 milliW) COMPASS beacon signals could be observed 
using, for example, 
the new VGOS VLBI system \cite{Haas-et-al-2015-a}, the Very Long Baseline Array (VLBA) \cite{Jones-et-al-2014-a}
or the proposed NG-VLA \cite{Lister-et-al-2018-a}. The goal is to provide nanoradian or better accuracy astrometry and meter-level or better transverse position accuracies with a relatively small number of observations per positioning fix. COMPASS would also support single frequency broadcast modes for simpler (and less accurate) navigation solutions. In either case COMPASS node would be assigned a unique identifier (a lunar equivalent of a Ethernet MAC address) which would used to multiplex the signal 
so that same beam interferometery could be performed with multiple beacons simultaneously in the same VLBI beam.

Table \ref{table:beam-widths} provides information on the possible COMPASS frequencies of operations and the resulting beamwidths at the highest operating frequencies for the terrestrial antennas involved. As the typical beamswidths are smaller than the Moon, same-beam observations will typically not be possible. Both VGOS and the NG-VLA have (or are planned to have) multiple collocated antennas, allowing for the simultaneous observations of COMPASS beacons and VLBI calibrators. 

COMPASS beacons carried by surface landers could serve as geodetic control points on the Lunar surface, directly tying the Lunar, planetary and extra-galactic reference frames, especially if the beacon was collocated with a suitable 
lunar laser ranging retroreflector 
\cite{Turyshev-et-al-2013-a}.
These measurements would meet a number of scientific goals; transverse VLBI measurements of lunar rotation would nicely complement the on-going program of radial measurements from Lunar Laser Ranging (LLR). VLBI beacon observations would 
also directly tie the lunar reference frame into the International Celestial Reference Frame \cite{Fey-et-al-2015-a} and would make it possible to extend the LLR reference frame to all surface or orbital operations.

\subsection{COMPASS UWB}

COMPASS will use Commercial-off-the-Shelf
(COTS) technology to the maximum extent possible. 
Crucial to the COMPASS accuracy goals is the production of coherent multi-frequency 
Ultra-WideBand (UWB) GHZ beacons \cite{Eubanks-2018-b} and we are investigating the use or modification of the IEEE 
802.15.4-a-2011 standard for this purpose. This offers 
sub-nanosecond pulses with a 3.9 MHz (or higher) pulse repetition rate in 500 Mhz or 1 GHz bandwidth channels. There is COTS industrial equipment available that supports centimeter level positioning and local communications networks with support for up to 11,000 communications nodes. Evaluation of this technology is on-going.

\section{Conclusions}

COMPASS will be a positioning system suitable for expanding the SSV into cislunar space based 
small and relatively inexpensive UWB beacons. 
COMPASS techniques and technology will be applicable to a wide variety of exploration missions and will enable research into fundamental physics, celestial mechanics and the rotational and orbital dynamics of celestial bodies in the solar system. COMPASS will support direct scientific advances (by, say, enabling improvements in the knowledge of the rotational dynamics of a celestial body), indirect scientific advances (by, say, enabling a swarm spacecraft mission, such as a radio array in deep space, that would not be possible or cost effective otherwise), and a furtherance of space exploration in general (by lowering the cost and complexity of small spacecraft navigation).

While COMPASS technology will be applicable to a wide variety of exploration missions and commercial endeavors, a combined COMPASS beacon and Lunar Laser Ranging (LLR) laser retroreflector on the Lunar surface would provide an immediate and strong scientific return, directly tying the Lunar reference frame into the International Celestial Reference Frame (ICRF) and providing independent estimates of Lunar orbit and librations, providing complementary data useful in tests of General Relativity and in determining the internal dynamics of the Moon. Surface beacons would also provide phase reference sources for orbiting beacons in support of time-critical Lunar operations, such as during landing navigation.



\begin{thebibliography}{10}

\bibitem{King-et-al-1976-a}
R.~W. {King}, I.~{Counselman}, C.~C., and I.~I. {Shapiro}, ``{Lunar dynamics
  and selenodesy: Results from analysis of VLBI and laser data},'' {\em J.
  Geophys. Res.}, vol.~81, pp.~6251--6256, Dec 1976.

\bibitem{Slade-et-al-1977-a}
M.~A. {Slade}, R.~A. {Preston}, A.~W. {Harris}, L.~J. {Skjerve}, and D.~J.
  {Spitzmesser}, ``{ALSEP-Quasar Differential VLBI},'' {\em Moon}, vol.~17,
  pp.~133--147, Oct 1977.

\bibitem{Pogrebenko-et-al-2004-a}
S.~V. {Pogrebenko}, L.~I. {Gurvits}, R.~M. {Campbell}, I.~M. {Avruch}, J.-P.
  {Lebreton}, and C.~G.~M. {van't Klooster}, ``{VLBI tracking of the Huygens
  probe in the atmosphere of Titan},'' in {\em Planetary Probe Atmospheric
  Entry and Descent Trajectory Analysis and Science} (A.~{Wilson}, ed.),
  vol.~544 of {\em ESA Special Publication}, pp.~197--204, Feb. 2004.

\bibitem{Jones-et-al-2014-a}
D.~L. {Jones}, W.~M. {Folkner}, R.~A. {Jacobson}, C.~S. {Jacobs}, V.~{Dhawan},
  J.~{Romney}, and E.~{Fomalont}, ``{Astrometry of Cassini With the Vlba to
  Improve the Saturn Ephemeris},'' {\em Astron. J.}, vol.~149, p.~28, Jan.
  2015.

\bibitem{Cimo-et-al-2015-a}
G.~{Cimo}, D.~{Duev}, G.~{Molera Calves}, T.~{Bocanegra Bohamon},
  S.~{Pogrebenko}, and L.~{Gurvits}, ``{Ground-based VLBI observations of
  orbiters and landers.},'' {\em European Planetary Science Congress 2015, held
  27 September - 2 October, 2015 in Nantes, France}, vol.~10,
  pp.~EPSC2015--212, Oct. 2015.

\bibitem{Duev-et-al-2016-a}
D.~A. {Duev}, S.~V. {Pogrebenko}, G.~{Cim{\`o}}, G.~{Molera Calv{\'e}s}, T.~M.
  {Bocanegra Baham{\'o}n}, L.~I. {Gurvits}, M.~M. {Kettenis}, J.~{Kania},
  V.~{Tudose}, P.~{Rosenblatt}, J.-C. {Marty}, V.~{Lainey}, P.~{de Vicente},
  J.~{Quick}, M.~{Nickola}, A.~{Neidhardt}, G.~{Kronschnabl}, C.~{Ploetz},
  R.~{Haas}, M.~{Lindqvist}, A.~{Orlati}, A.~V. {Ipatov}, M.~A. {Kharinov},
  A.~G. {Mikhailov}, J.~E.~J. {Lovell}, J.~N. {McCallum}, J.~{Stevens}, S.~A.
  {Gulyaev}, T.~{Natush}, S.~{Weston}, W.~H. {Wang}, B.~{Xia}, W.~J. {Yang},
  L.-F. {Hao}, J.~{Kallunki}, and O.~{Witasse}, ``{Planetary Radio
  Interferometry and Doppler Experiment (PRIDE) technique: A test case of the
  Mars Express Phobos fly-by},'' {\em Astron. Astrophys.}, vol.~593, p.~A34,
  Sept. 2016.

\bibitem{Goossens-et-al-2011-a}
S.~{Goossens}, K.~{Matsumoto}, Q.~{Liu}, F.~{Kikuchi}, K.~{Sato}, H.~{Hanada},
  Y.~{Ishihara}, H.~{Noda}, N.~{Kawano}, N.~{Namiki}, T.~{Iwata}, F.~G.
  {Lemoine}, D.~D. {Rowlands}, Y.~{Harada}, and M.~{Chen}, ``{Lunar gravity
  field determination using SELENE same-beam differential VLBI tracking
  data},'' {\em Journal of Geodesy}, vol.~85, pp.~205--228, Apr 2011.

\bibitem{Kikuchi-et-al-2009-a}
F.~{Kikuchi}, Q.~{Liu}, H.~{Hanada}, N.~{Kawano}, K.~{Matsumoto}, T.~{Iwata},
  S.~{Goossens}, K.~{Asari}, Y.~{Ishihara}, S.~{Tsuruta}, T.~{Ishikawa},
  H.~{Noda}, N.~{Namiki}, N.~{Petrova}, Y.~{Harada}, J.~{Ping}, and
  S.~{Sasaki}, ``{Picosecond accuracy VLBI of the two subsatellites of SELENE
  (KAGUYA) using multifrequency and same beam methods},'' {\em Radio Science},
  vol.~44, p.~RS2008, Mar 2009.

\bibitem{Klopotek-et-al-2019-a}
G.~{Klopotek}, T.~{Hobiger}, R.~{Haas}, F.~{Jaron}, L.~{La Porta},
  A.~{Nothnagel}, Z.~{Zhang}, S.~{Han}, A.~{Neidhardt}, and C.~{Pl{\"o}tz},
  ``{Position determination of the Chang'e 3 lander with geodetic VLBI},'' {\em
  Earth, Planets, and Space}, vol.~71, p.~23, Feb 2019.

\bibitem{Liu-et-al-2014-a}
Q.~{Liu}, X.~{Zheng}, Y.~{Huang}, P.~{Li}, Q.~{He}, Y.~{Wu}, L.~{Guo}, and
  M.~{Tang}, ``{Monitoring motion and measuring relative position of the
  Chang'E-3 rover},'' {\em Radio Science}, vol.~49, pp.~1080--1086, Nov 2014.

\bibitem{Yan-et-al-2011-a}
J.~{Yan}, F.~{Li}, Q.~{Liu}, J.~{Ping}, Z.~{Zhong}, and J.~{Li}, ``{Proposal of
  application of same beam VLBI measurements in precision orbit determination
  of lunar orbiter and return capsule and lunar gravity field simulation in
  Chang'E-3 mission},'' {\em Advances in Space Research}, vol.~48,
  pp.~1676--1681, Nov 2011.

\bibitem{Zhang-et-al-2017-b}
Z.~{Zhang}, S.~{Han}, A.~{Nothnagel}, L.~{La Porta}, S.~{Halsig}, A.~{Iddink},
  F.~{Jaron}, R.~{Haas}, J.~{Lovell}, and A.~{Neidhardt}, ``{Initial
  Estimations of the Lunar Lander Position by OCEL Observations},'' in {\em
  23rd European VLBI Group for Geodesy and Astrometry Working Meeting},
  vol.~23, pp.~200--204, Nov 2017.

\bibitem{Bauer-et-al-2006-a}
F.~H. {Bauer}, M.~Moreau, M.~Dahle-Melsaether, W.~Petrofski, B.~Stanton,
  S.~Thomason, G.~Harris, R.~Sena, and L.~{Parker~Temple~III}, ``{The GPS Space
  Service Volume},'' in {\em ION GNSS 2006}, ION GNSS, pp.~2503 -- 2514, Sept.
  2006.

\bibitem{Ashman-et-al-2018-a}
B.~Ashman, F.~H. Bauer, J.~Parker, and J.~Donaldson, {\em GPS Operations in
  High Earth Orbit: Recent Experiences and Future Opportunities}, pp.~AIAA
  2018--2568.
\newblock American Institute of Aeronautics and Astronautics, 2018.

\bibitem{Eubanks-et-al-2017-a}
T.~M. {Eubanks}, T.~{Cash}, B.~{Blair}, and M.~E. {Eubanks}, ``{The
  Femtospacecraft Asteroid Impact Mission (FAIM): A Low Cost Mission to Monitor
  the DART Impact on the Didymoon},'' in {\em Lunar and Planetary Science
  Conference}, vol.~48 of {\em Lunar and Planetary Science Conference},
  p.~1577, Mar. 2017.

\bibitem{Haas-et-al-2015-a}
R.~{Haas}, A.~{Nothnagel}, and B.~{Petrachenko}, ``{VGOS - the VLBI Global
  Observing System of the IVS},'' {\em IAU General Assembly}, vol.~22,
  p.~2257511, Aug. 2015.

\bibitem{Lister-et-al-2018-a}
M.~L. {Lister}, K.~I. {Kellermann}, and P.~{Kharb}, {\em {High-resolution
  Imaging of Radio Jets Launched by Active Galactic Nuclei: New Insights on
  Formation, Structure, and Evolution Enabled by the ngVLA}}, vol.~517 of {\em
  Astronomical Society of the Pacific Conference Series}, p.~619.
\newblock Astronomical Society of the Pacific, 2018.

\bibitem{Turyshev-et-al-2013-a}
S.~G. {Turyshev}, J.~G. {Williams}, W.~M. {Folkner}, G.~M. {Gutt}, R.~T.
  {Baran}, R.~C. {Hein}, R.~P. {Somawardhana}, J.~A. {Lipa}, and S.~{Wang},
  ``{Corner-cube retro-reflector instrument for advanced lunar laser
  ranging},'' {\em Experimental Astronomy}, vol.~36, pp.~105--135, Aug. 2013.

\bibitem{Fey-et-al-2015-a}
A.~L. {Fey}, D.~{Gordon}, C.~S. {Jacobs}, C.~{Ma}, R.~A. {Gaume}, E.~F.
  {Arias}, G.~{Bianco}, D.~A. {Boboltz}, S.~{B{\"o}ckmann}, S.~{Bolotin},
  P.~{Charlot}, A.~{Collioud}, G.~{Engelhardt}, J.~{Gipson}, A.-M. {Gontier},
  R.~{Heinkelmann}, S.~{Kurdubov}, S.~{Lambert}, S.~{Lytvyn}, D.~S.
  {MacMillan}, Z.~{Malkin}, A.~{Nothnagel}, R.~{Ojha}, E.~{Skurikhina},
  J.~{Sokolova}, J.~{Souchay}, O.~J. {Sovers}, V.~{Tesmer}, O.~{Titov},
  G.~{Wang}, and V.~{Zharov}, ``{The Second Realization of the International
  Celestial Reference Frame by Very Long Baseline Interferometry},'' {\em
  Astron. J.}, vol.~150, p.~58, Aug. 2015.

\bibitem{Eubanks-2018-b}
T.~M. {Eubanks}, ``{MilliWatt Lunar VLBI Beacons: Surviving the Lunar Night},''
  in {\em Survive and Operate Through the Lunar Night Workshop}, vol.~2106,
  p.~7027, Nov 2018.

\end{thebibliography}
\bibliographystyle{ieeetr}

\end{document}